\documentclass{ws-ijmpa}
\usepackage[super,compress]{cite}
\usepackage{graphicx}

\def\Bsmumu{B_s\to\mu^+\mu^-}

\newcommand{\gev}{\, {\rm GeV}}
\newcommand{\mev}{\, {\rm MeV}}

\begin{document}
\markboth{Robert Fleischer}
{Probing New Physics with $B^0_s\to\mu^+\mu^-$: Status and Perspectives}

%
\catchline{}{}{}{}{}
%

\title{PROBING NEW PHYSICS WITH $B^0_s\to\mu^+\mu^-$:\\ STATUS AND PERSPECTIVES}

\author{ROBERT FLEISCHER}

\address{Nikhef, Science Park 105, NL-1098 XG Amsterdam, Netherlands\\
Department of Physics and Astronomy, Vrije Universiteit Amsterdam,\\ NL-1081 HV Amsterdam, Netherlands\\
Robert.Fleischer@nikhef.nl
}

\maketitle

\begin{history}
\end{history}

\begin{abstract}
The rare decay $B^0_s \to \mu^+\mu^-$ plays a key role for the testing of the Standard Model. 
It is pointed out that the sizable decay width difference $\Delta\Gamma_s$ of the $B_s$-meson 
system affects this channel in a subtle way. As a consequence, its calculated Standard Model 
branching ratio has to be upscaled by about $10\%$. Moreover, the sizable $\Delta\Gamma_s$ 
makes a new observable through the effective $B^0_s\to \mu^+\mu^-$ lifetime accessible, 
which probes New Physics in a way complementary to the branching ratio and adds an 
exciting new topic to the agenda for the high-luminosity upgrade of the LHC. Further probes 
of New Physics are offered by a CP-violating rate asymmetry. Correlations between these 
observables and the $B^0_s \to \mu^+\mu^-$ branching ratio are illustrated for specific 
models of New Physics. 
\end{abstract}

\ccode{PACS numbers: 13.20.He, 12.60.-i}


%
%
%
\section{Introduction}\label{sec:intro}	
\begin{figure}[t]
\centerline{\includegraphics[width=8.5cm]{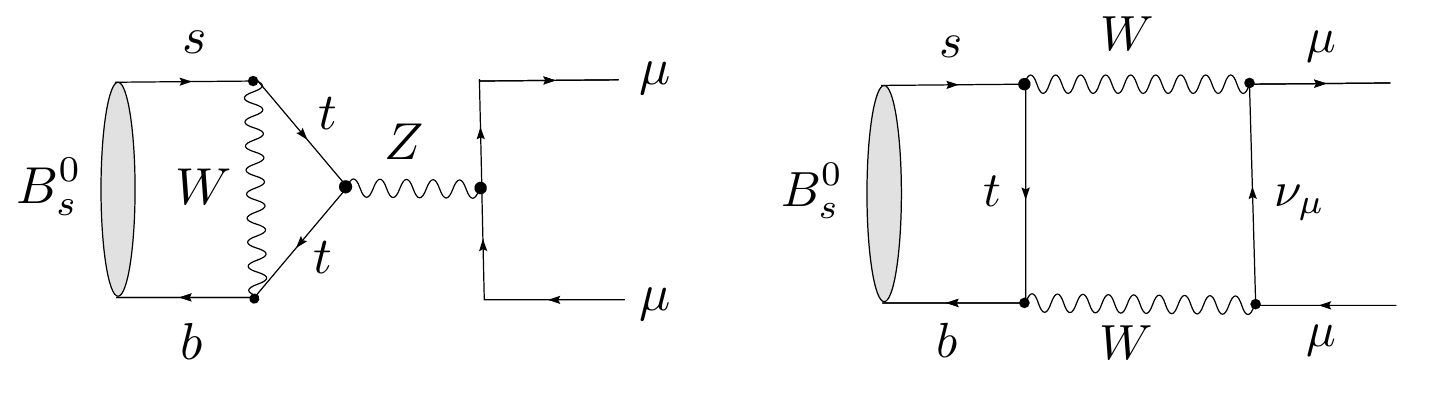}}
\caption{Penguin and box diagrams contributing to the $B^0_s\to\mu^+\mu^-$ decay
in the Standard Model.\label{fig-diag}}
\end{figure}

In the Standard Model (SM), the decay $B^0_s\to\mu^+\mu^-$ arises only from loop contributions
related to penguin and box topologies, as can be seen in Fig.~\ref{fig-diag}, 
and is helicity suppressed, resulting in a strongly suppressed
branching ratio which is proportional to $m_\mu^2$. As only leptons are present in the final state, 
the hadronic sector is very simple and described by a single non-perturbative parameter, 
the $B_s$ decay constant $F_{B_{s}}$, which is defined through the relation
\begin{equation}\label{FBs-def}
\langle 0| \bar b \gamma_5\gamma_\mu s | B^0_s(p)\rangle = i F_{B_s} p_\mu .
\end{equation}
In view of these features, $B^0_s\to\mu^+\mu^-$ belongs to the cleanest rare $B$ 
decays Nature has to offer and represents an outstanding probe for physics beyond the SM.

In the SM, the parametric dependence on the relevant input parameters is given as follows
\cite{BGGI,BFGK}:
\begin{eqnarray}
\lefteqn{{\rm BR}(\Bsmumu)_{\rm SM} = 3.25\times 10^{-9}}\nonumber\\
&& \times \left[\frac{M_t}{173.2 \gev}\right]^{3.07}\left[\frac{F_{B_s}}{225\mev}\right]^2
\left[\frac{\tau_{B_s}}{1.500 {\rm ps}}\right]\left|\frac{V_{tb}^*V_{ts}}{0.0405}\right|^2,\label{BR-par}
\end{eqnarray}
where $M_t$ is the top-quark mass, $\tau_{B_s}$ the $B^0_s$-meson lifetime, and 
$V_{tb}^*V_{ts}$ the relevant combination of elements of the Cabibbo--Kobayashi--Maskawa 
(CKM) matrix.

\begin{figure}[b]
\vspace*{-0.5truecm}
\centerline{\includegraphics[width=4.7cm]{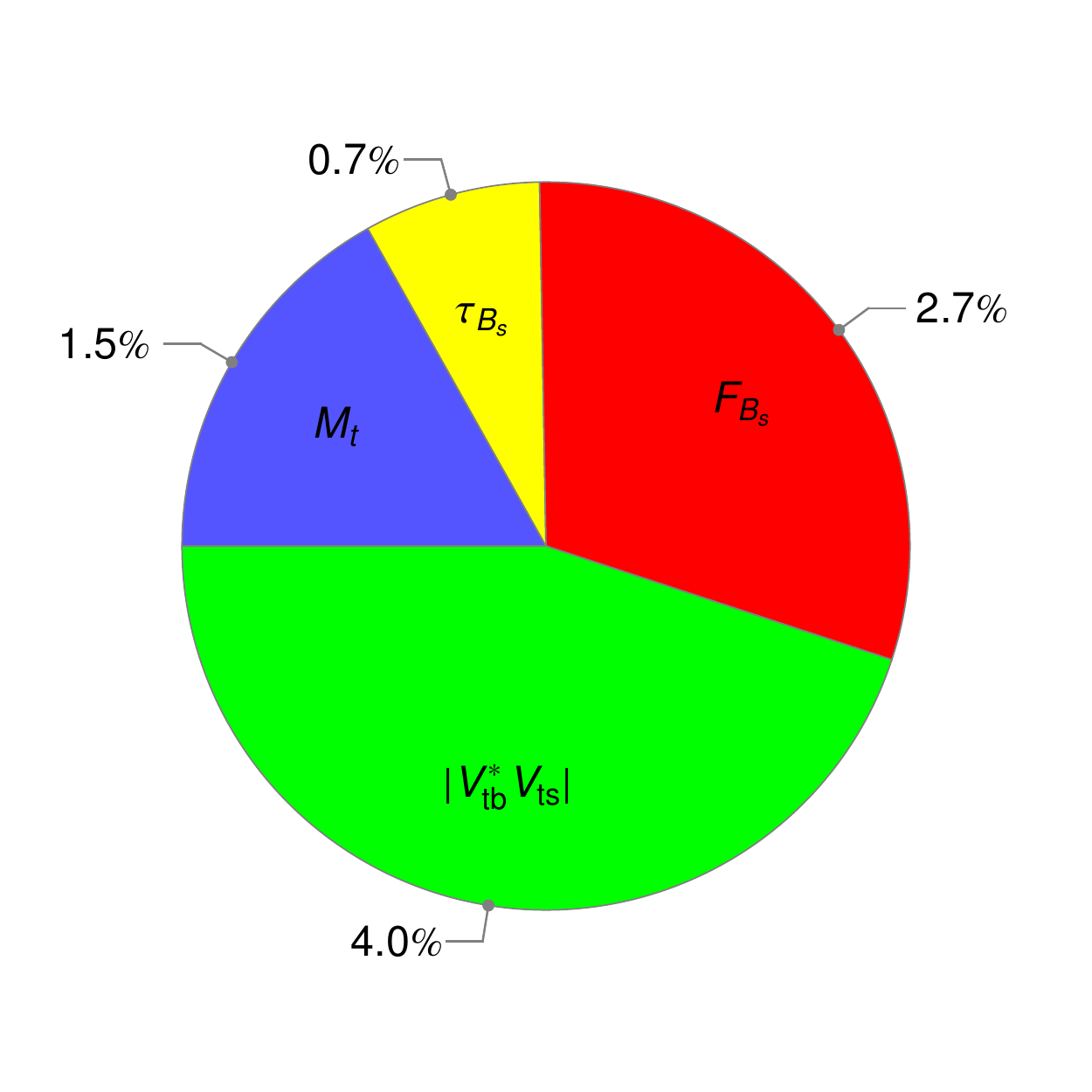}}
\vspace*{-0.5truecm}
\caption{The error budget of ${\rm BR}(\Bsmumu)_{\rm SM}$ related to the 
various input parameters.\cite{BFGK}.\label{fig:budget}}
\end{figure}

Concerning the SM prediction of the $B^0_s\to\mu^+\mu^-$  branching ratio, there has recently
been important progress in lattice QCD,\cite{Lattice} which is reflected by the result
$F_{B_s}=(227.7\pm 4.5)\,\mev$, while progress on the experimental side\cite{HFAG} led to
 an improved measurement of $\tau_{B_s}=(1.516 \pm 0.011)\,{\rm ps}$. In 
Fig.~\ref{fig:budget}, the corresponding error budget for the SM value of the $B^0_s\to\mu^+\mu^-$
branching ratio is shown. On the theoretical side,\cite{Theo-1} there was important progress thanks 
to a very impressive calculation of NLO electroweak effects\cite{Theo-2} and NNLO QCD 
matching corrections \cite{Theo-3}, resulting in 
\begin{equation}
{\rm BR}(\Bsmumu)_{\rm SM} = (3.38\pm 0.22)\times 10^{-9},
\end{equation}
which supersedes the prediction in Ref.~\refcite{BFGK}.

Thanks to the impact of New Physics (NP), i.e.\ physics beyond the SM, the branching ratios
of the $B^0_{s,d}\to\mu^+\mu^-$ decays could have been enhanced significantly, in particular
in supersymmetric flavor models (see, for instance, Refs.~\refcite{straub,BuGir-12} and references
therein). In view of this feature, there was the exciting possibility to observe $B^0_s\to\mu^+\mu^-$
already at the Tevatron in the previous decade. As the Tevatron collider is now legacy, the 
CDF and D0 collaborations have presented their final results 
on the search for $B^0_s\to\mu^+\mu^-$, corresponding to the 95\% C.L. upper bounds 
$15 \times 10^{-9}$ and $31 \times 10^{-9}$, respectively.\cite{CDF-bound,D0-bound}

The analysis of the rare $B^0_s\to\mu^+\mu^-$ channel 
is now conducted at the Large Hadron Collider (LHC). 
The ATLAS experiment\cite{ATLAS-mumu} has set the upper bound 
$\mbox{BR}(B_s \to \mu^+\mu^-) < 15\times 10^{-9}$ \,(95\% C.L.), while the {\it first evidence} 
for $B^0_s\to\mu^+\mu^-$ was reported by the CMS\cite{CMS-mumu} and 
LHCb\cite{LHCb-mumu} collaborations in 2013, with the results 
$\mbox{BR}(B_s \to \mu^+\mu^-) = (3.0^{+1.0}_{-0.9}) \times 10^{-9}$  and 
$(2.9^{+1.1}_{-1.0}) \times 10^{-9}$, respectively. The average of these LHC measurements
is given as follows:\cite{LHC-Bsmumu-average}
\begin{equation}
\mbox{BR}(B_s \to \mu^+\mu^-) = (2.9 \pm 0.7) \times 10^{-9}.
\end{equation}
It should be noted that the limiting factor for the 
$\mbox{BR}(B_s\to\mu^+\mu^-)$ measurement -- and actually all $B_s$ branching ratios -- 
is given by the ratio $f_s/f_{d}$ of the corresponding fragmentation functions.\cite{FST,FF-lat}

It will be interesting to keep an eye on $B^0_d\to\mu^+\mu^-$. The current
information on the branching ratio reported by the CMS and LHCb collaborations is given by
\begin{equation}
\mbox{BR}(B_d \to \mu^+\mu^-)=\left\{
\begin{array}{ll}
(3.5^{+2.1}_{-1.8}) \times 10^{-10}  < 11 \times 10^{-10} & \mbox{(CMS)}\\
(3.7^{+2.4}_{-2.1}) \times 10^{-10} < 7.4 \times 10^{-10} & \mbox{(LHCb),}
\end{array}
\right.
\end{equation}
where the upper bounds refer to the 95\% C.L., resulting in the LHC
average\cite{LHC-Bsmumu-average}
\begin{equation}\label{Bd-BR}
\mbox{BR}(B_d \to \mu^+\mu^-) = (3.6^{+1.6}_{-1.4}) \times 10^{-10}.
\end{equation}
On the other hand, the SM prediction is given as follows:\cite{Theo-1}
\begin{equation}
\mbox{BR}(B_d \to \mu^+\mu^-)_{\rm SM} = (1.06\pm0.09) \times 10^{-10}.
\end{equation}
The current experimental errors are too big to draw conclusions about physics beyond the SM
model. However, should a result around the central value in (\ref{Bd-BR}) be established in the 
future, it would immediately rule out the SM and its extensions with ``Minimal Flavor 
Violation" (MFV).

\section{Branching Ratios of $B_s$ Decays for $\Delta\Gamma_s\not=0$}	
Thanks to $B_s^0$--$\bar B^0_s$ mixing, an initially, i.e.\ at time $t=0$, present $B^0_s$
meson evolves into a time-dependent linear combination of $|B^0_s\rangle$ and
$|\bar B^0_s\rangle$ states:\cite{RF-rev}
\begin{equation}
|B_s(t)\rangle=a(t)|B^0_s\rangle+ b(t)|\bar B^0_s\rangle.
\end{equation}
The time evolution is described by an appropriate Sch\"ordinger equation. It is solved
by introducing mass eigenstates $B_s^{\rm H}$ (``heavy") and $B_s^{\rm L}$ (``light") which
are characterised by the differences
\begin{equation}
\Delta M_s\equiv M_{\rm H}^{(s)}-M_{\rm L}^{(s)} \quad\mbox{and}\quad
\Delta\Gamma_s\equiv\Gamma_{\rm L}^{(s)}-\Gamma_{\rm H}^{(s)}
\end{equation}
of their masses and decay widths, respectively. A characteristic feature of the $B_s$-meson system
is a sizeable decay width difference $\Delta\Gamma_s$, which has been expected on theoretical
grounds since decades; for a recent review, see Ref.~\refcite{Lenz}. In contrast, 
the decay width difference of the 
$B_d$-meson system is negligibly small. Recently, a non-zero value of $\Delta\Gamma_s$ 
has actually been established at the $6\,\sigma$ level by LHCb:\cite{LHCb-DGs}
\begin{equation}\label{ys-range}
        y_s \equiv \frac{\Delta\Gamma_s}{2\,\Gamma_s}\equiv
        \frac{\Gamma_{\rm L}^{(s)} - \Gamma_{\rm H}^{(s)}}{2\,\Gamma_s}= 0.075 \pm 0.012,
\end{equation}
where the decay width parameter $y_s$ characterises the impact of $\Delta\Gamma_s$ in
formulae to be given below.

\begin{figure}
\centerline{\includegraphics[width=7.8cm]{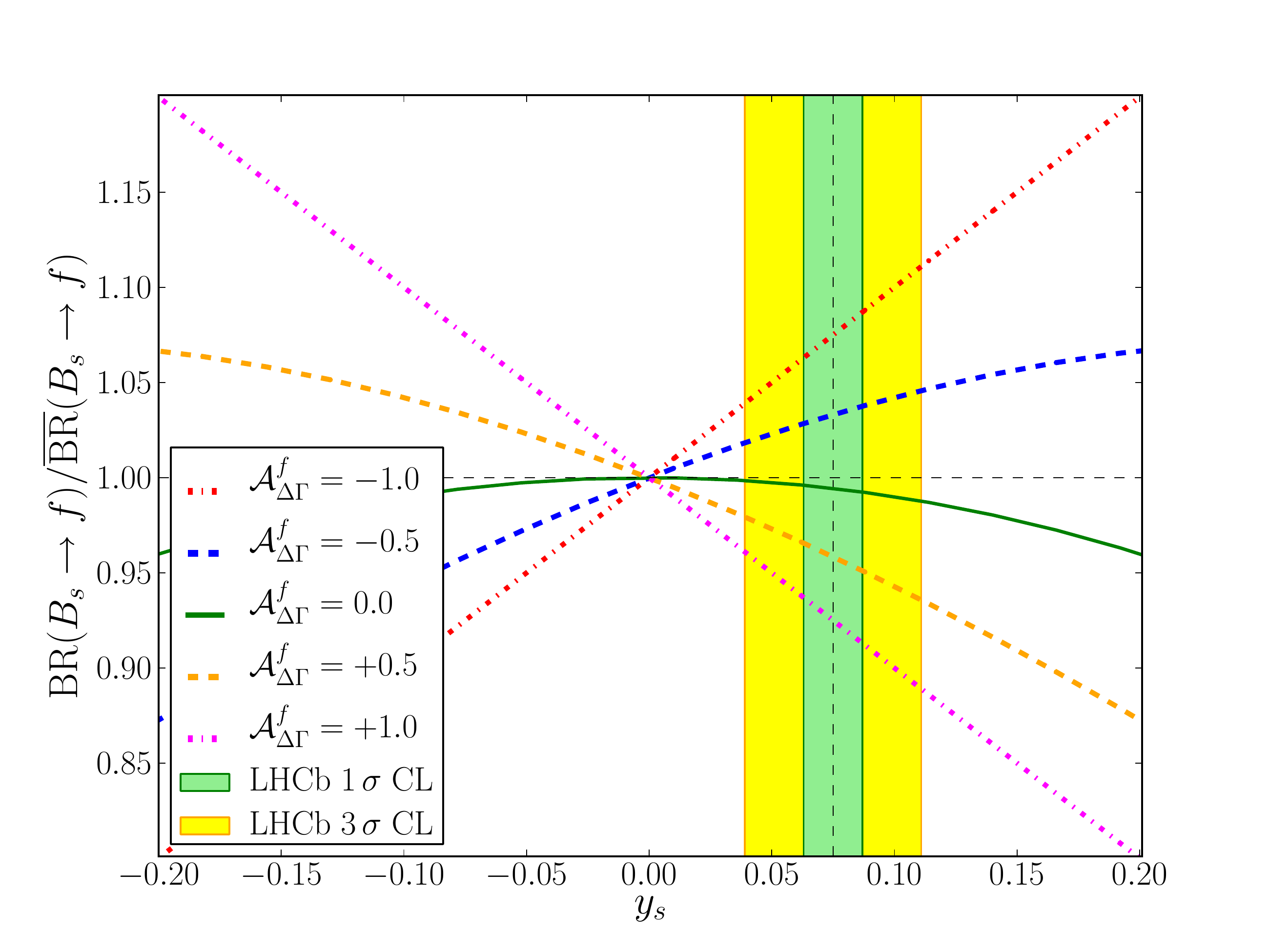}}
\caption{The ratio of the theoretical to the experimental branching
ratio of $B_s\to f$ as a function of the decay width parameter $y_s$ for various 
values of the observable ${\cal A}^f_{\Delta\Gamma}$.\label{fig-BR-rat}}
\end{figure}

In view of the sizeable decay width difference $\Delta\Gamma_s$ special care 
has to be taken when dealing with branching ratios of $B_s$-meson 
decays.\cite{BR-paper,Bsmumu-paper}
The starting point for the analysis of branching ratios is the following ``untagged" rate, where
no distinction between initially, i.e.\ at time $t=0$, present $B^0_s$ or $\bar B^0_s$ mesons
is made:
\begin{displaymath}
        \langle \Gamma(B_s(t)\to f)\rangle
        \equiv\ \Gamma(B^0_s(t)\to f)+ \Gamma(\bar B^0_s(t)\to f) =
        R^f_{\rm H} e^{-\Gamma_{\rm H}^{(s)} t} + R^f_{\rm L} e^{-\Gamma_{\rm L}^{(s)} t}
\end{displaymath}
\begin{equation}
        = \left(R^f_{\rm H} + R^f_{\rm L}\right) e^{-\Gamma_s\,t}
        \left[ \cosh\left(\frac{y_s\, t}{\tau_{B_s}}\right)+
        {\cal A}^f_{\rm \Delta\Gamma}\,\sinh\left(\frac{y_s\, t}
        {\tau_{B_s}}\right)\right].
\end{equation}
The ``experimental" branching ratio refers to the time-integrated untagged rate:\cite{DDF}
\begin{displaymath}
\hspace*{-2.0truecm}{\rm BR}\left(B_s \to f\right)_{\rm exp} \equiv
\overline{\rm BR}\left(B_s \to f\right)
        \equiv \frac{1}{2}\int_0^\infty \langle \Gamma(B_s(t)\to f)\rangle\, dt 
\end{displaymath}
\begin{equation}
= \frac{1}{2}\left[ \frac{R^f_{\rm H}}{\Gamma^{(s)}_{\rm H}} + 
        \frac{R^f_{\rm L}}{\Gamma^{(s)}_{\rm L}}\right]
        = \frac{\tau_{B_s}}{2}\left(R^f_{\rm H} + R^f_{\rm L}\right)
        \left[\frac{1 + {\cal A}^f_{\Delta\Gamma}\, y_s}{1-y_s^2} \right].
\end{equation}
On the other hand, in theoretical analyses, usually the following ``theoretical" branching ratio
is considered:\cite{RF-JpsiK,RF-KK,FFM}
\begin{equation}
{\rm BR}\left(B_s \to f\right)
\equiv 
        \frac{\tau_{B_s}}{2}\langle \Gamma(B^0_s(t)\to f)\rangle\Big|_{t=0}
        = \frac{\tau_{B_s}}{2}\left(R^f_{\rm H} + R^f_{\rm L}\right).
\end{equation}
By considering $t=0$, the effect of $B^0_s$--$\bar B^0_s$ mixing is ``switched off". 
The advantage of this definition is that it allows a straightforward comparison with 
the branching ratios of $B^0_d$ or $B_u^+$ mesons by means of the $SU(3)$
flavor symmetry. Using the relation
\begin{equation}
        {\rm BR}\left(B_s \to f\right)
        = \left[\frac{1-y_s^2}{1 + {\cal A}^f_{\Delta\Gamma}\, y_s}\right]
        \overline{\rm BR}\left(B_s \to f\right),
\end{equation}
the experimental branching ratio can be converted into the theoretical branching ratio.\cite{BR-paper}
While the decay width parameter $y_s$ has already
been measured, the observable ${\cal A}^f_{\Delta\Gamma}$ depends on the considered decay 
and generally involves non-perturbative parameters. As can be seen in Fig.~\ref{fig-BR-rat}, 
the differences between the two branching ratio concepts can be as large as 
${\cal O}(10\%)$ for the measured value of $y_s$ in Eq.~(\ref{ys-range}).

In order to determine the process-dependent value of ${\cal A}^f_{\Delta\Gamma}$, typically
theoretical assumptions have to be made, such as using the $SU(3)$ flavor symmetry in the
case of non-leptonic $B_s$ decays (for a compilation of results, see Ref.~\refcite{BR-paper}).
For the extraction of the theoretical branching ratio, it is desirable to avoid theoretical input. 
This can be achieved by means of a measurement of the effective $B_s\to f$ decay 
lifetime:\cite{BR-paper}
\begin{equation}\label{tau-eff}
\tau_f \equiv \frac{\int_0^\infty t\,\langle \Gamma(B_s(t)\to f)\rangle\, dt}
        {\int_0^\infty \langle \Gamma(B_s(t)\to f)\rangle\, dt}= \frac{\tau_{B_s}}{1-y_s^2}\left[\frac{1+2\,{\cal A}^f_{\Delta\Gamma}y_s + y_s^2}
        {1 + {\cal A}^f_{\Delta\Gamma} y_s}\right],
\end{equation}
which yields
\begin{equation}\label{BR-conv}
{\rm BR}\left(B_s \to f\right)
= \left[2 - \left(1-y_s^2\right)\frac{\tau_f}{\tau_{B_s}}\right]\overline{\rm BR}\left(B_s \to f\right).
\end{equation}
On the right-hand side of this expression, only quantities enter which can be measured. Once
information on the effective decay lifetime is available, which requires a time-dependent 
measurement of the untagged $B_s$ decay rate as can be seen in Eq.~(\ref{tau-eff}), 
Eq.~(\ref{BR-conv}) is advocated for
the determination of theoretical branching ratios for particle listings by the Particle Data 
Group.\cite{BR-paper}  For a discussion of the branching ratio measurements 
of $B_s\to VV$ decays into two vector mesons in the presence of a sizeable value of
$\Delta\Gamma_s$, such as $B_s\to J/\psi \phi$, $B_s\to K^{*0}\bar K^{*0}$  and 
$B_s\to D_s^{*+}D_s^{*-}$, the reader is referred to 
Refs.~\refcite{BR-paper,LHCb-BsKastKast,DGMV}.

\section{The $B^0_s\to\mu^+\mu^-$ Observables}	
The subtleties discussed in the previous section apply also to the decay $B^0_s\to\mu^+\mu^-$. 
In this case, the branching ratio serves as a sensitive probe for New Physics and it is 
essential -- in view of the current experimental situation with branching ratio measurements
falling into the SM regime -- to assess the impact of the sizeable decay width difference
$\Delta\Gamma_s$. As we will see below, this quantity offers -- apart from the complication for
the analysis of the branching ratio -- a new observable to search for New Physics. 
The discussion in this section follows closely Ref.~\refcite{Bsmumu-paper}.

\subsection{Decay Amplitude}
The low-energy effective Hamiltonian describing the decay $\bar B^0_s\to\mu^+\mu^-$
is given as follows:
\begin{equation}\label{Heff}
{\cal H}_{\rm eff}=-\frac{G_{\rm F}}{\sqrt{2}\pi} V_{ts}^\ast V_{tb} \alpha
\bigl[C_{10} O_{10} + C_{S} O_S + C_P O_P
+ C_{10}' O_{10}' + C_{S}' O_S' + C_P' O_P' \bigr],
\end{equation}
where $G_{\rm F}$ is Fermi's constant and $\alpha$ denotes the QED fine structure constant. 
In the general Hamiltonian in (\ref{Heff}), only four-fermion operators with non-vanishing 
$\bar B^0_s\to \mu^+\mu^-$ matrix elements are included. They take the following form:
\begin{equation}
\begin{array}{rclrcl}
O_{10}&=&(\bar s \gamma_\mu P_L b) (\bar\ell\gamma^\mu \gamma_5\ell), &
\hspace*{0.5truecm} O_{10}'&=&(\bar s \gamma_\mu P_R b) (\bar\ell\gamma^\mu \gamma_5\ell) \\
O_S&=&m_b (\bar s P_R b)(\bar \ell \ell), &
O_S'&=&m_b (\bar s P_L b)(\bar \ell \ell) \\
O_P&=&m_b (\bar s P_R b)(\bar \ell \gamma_5 \ell), &
O_P'&=&m_b (\bar s P_L b)(\bar \ell \gamma_5 \ell),
\end{array}
\end{equation}
where $P_{L,R}\equiv(1\mp\gamma_5)/2$ and $m_b$ denotes the $b$-quark mass.
The Wilson coefficients $C_i$,  $C_i'$  encode the short-distance physics. In the SM, 
only the operator $O_{10}$ contributes with a {\it real} coefficient $C_{10}^{\rm SM}$. 
The outstanding 
feature of the $\bar B^0_s\to \mu^+\mu^-$ channel is the sensitivity to (pseudo-)scalar 
lepton densities, which are described by the $O_{(P)S}$, $O_{(P)S}'$ operators, having
Wilson coefficients which are still largely unconstrained.\cite{AS,WA}

In order to calculate the decay amplitude, it is convenient to go to the rest frame of the
decaying $\bar B^0_s$ meson and to distinguish between the
$\mu^+_{\rm L}\mu^-_{\rm L}$ and $\mu^+_{\rm R}\mu^-_{\rm R}$ helicity configurations
which are related to each other through a CP transformation:
\begin{equation}
|(\mu_{\rm L}^+\mu_{\rm L}^-)_{\rm CP}\rangle \equiv ({\cal CP})|\mu_{\rm L}^+\mu_{\rm L}^-\rangle=
e^{i\phi_{\rm CP}(\mu\mu)}|\mu_{\rm R}^+\mu_{\rm R}^-\rangle.
\end{equation}
The $e^{i\phi_{\rm CP}(\mu\mu)}$ is a convention-dependent phase factor which cancels 
in the observables discussed below. The general expression for the decay amplitude (with
$\eta_{\rm L}=+1$ and  $\eta_{\rm R}=-1$) reads as
\begin{eqnarray}
\lefteqn{A(\bar B^0_s \to \mu_\lambda^+\mu_\lambda^-)=\langle \mu_\lambda^-\mu_\lambda^+|
{\cal H}_{\rm eff}| \bar B^0_s \rangle = }
\nonumber\\
&& -\frac{G_{\rm F}}{\sqrt{2}\pi} V_{ts}^\ast V_{tb} \alpha  F_{B_s} M_{B_s}   m_\mu
C_{10}^{\rm SM} e^{i\phi_{\rm CP}(\mu\mu)(1-\eta_\lambda)/2}
\left [\eta_\lambda P +  S    \right].\label{Ampl}
\end{eqnarray}
Here the following combinations of Wilson coefficient functions were introduced:
\begin{equation}\label{P-expr}
P\equiv |P|e^{i\varphi_P}\equiv \frac{C_{10}-C_{10}'}{C_{10}^{\rm SM}}+
{\frac{M_{B_s} ^2}{2\, m_\mu}
\left(\frac{m_b}{m_b+m_s}\right)\left(\frac{C_P-C_P'}{C_{10}^{\rm SM}}\right)}
\,\stackrel{\rm SM}{\longrightarrow}\, 1
\end{equation}
\begin{equation}\label{S-expr}
S\equiv |S|e^{i\varphi_S}\equiv \sqrt{1-4\frac{m_\mu^2}{M_{B_s}^2}}
{\frac{M_{B_s} ^2}{2\, m_\mu}\left(\frac{m_b}{m_b+m_s}\right)
\left(\frac{C_S-C_S'}{C_{10}^{\rm SM}}\right)}
\,\stackrel{\rm SM}{\longrightarrow}\, 0,
\end{equation}
where the $\varphi_{P,S}$ are CP-violating NP phases. As indicated, the $P$ and $S$ 
were introduced in such a way that they equal $1$ and $0$ in the SM case, respectively. 
In Eq.~(\ref{Ampl}), $F_{B_s}$ denotes the $B_s$ decay constant as 
introduced in Eq.~(\ref{FBs-def}), $M_{B_s}$ and $m_\mu$ are the $B_s$ and muon masses, 
respectively, while $m_s$ denotes the strange-quark mass.

\subsection{CP Asymmetries}
For the calculation of the CP asymmetries and the untagged rate of $B_s\to\mu^+\mu^-$, 
the following observable is required:\cite{RF-rev}
\begin{equation}
\xi_\lambda\equiv - e^{-i\phi_s}\left[ e^{i\phi_{\rm CP}(B_s)}
\frac{A(\bar B^0_s\to \mu_\lambda^+\mu_\lambda^-)}{A(B^0_s\to \mu_\lambda^+\mu_\lambda^-)}
\right],
\end{equation}
which involves also the amplitude
\begin{equation}
A(B^0_s \to \mu_\lambda^+\mu_\lambda^-)=
\langle \mu_\lambda^-\mu_\lambda^+|{\cal H}_{\rm eff}^\dagger | B^0_s \rangle.
\end{equation}
Using  $({\cal CP})^\dagger ({\cal CP})=\hat 1$
and $({\cal CP})|B^0_s\rangle=e^{i\phi_{\rm CP}(B_s)}|\bar B^0_s\rangle$ yields the expression
\begin{eqnarray}
\lefteqn{A(B^0_s \to \mu_\lambda^+\mu_\lambda^-)= -\frac{G_{\rm F}}{\sqrt{2}\pi} 
V_{ts} V_{tb}^\ast \alpha f_{B_s} M_{B_s}  m_\mu C_{10}^{\rm SM}}\nonumber\\
&&\times e^{i[\phi_{\rm CP}(B_s)+\phi_{\rm CP}(\mu\mu)(1-\eta_\lambda)/2]}
\left [-\eta_\lambda P^\ast +  S^\ast    \right].
\end{eqnarray}
The convention-dependent phases cancel in $\xi_\lambda$, which takes the form
\begin{equation}
\xi_\lambda =-\left[\frac{+\eta_\lambda P \,+\,  S}{-\eta_\lambda P^\ast + S^\ast }\right],
\end{equation}
satisfying the relation
\begin{equation}
\xi_{\rm L}^{\phantom{\ast}} \xi_{\rm R}^{\ast} = 
\xi_{\rm R}^{\phantom{\ast}} \xi_{\rm L}^{\ast} =1.
\end{equation}

The time-dependent CP asymmetry, which requires {\it tagging} to distinguish initially, i.e.\ at
time $t=0$, present $B^0_s$ and $\bar B^0_s$ mesons, is given by
\begin{equation}\label{TD-CP}
\frac{\Gamma(B^0_s(t)\to \mu_\lambda^+\mu^-_\lambda)-
\Gamma(\bar B^0_s(t)\to \mu_\lambda^+
\mu^-_\lambda)}{\Gamma(B^0_s(t)\to \mu_\lambda^+\mu^-_\lambda)+
\Gamma(\bar B^0_s(t)\to \mu_\lambda^+\mu^-_\lambda)}
=\frac{C_\lambda\cos(\Delta M_st)+S_\lambda\sin(\Delta M_st)}{\cosh(y_st/\tau_{B_s}) + 
{\cal A}_{\Delta\Gamma}^\lambda \sinh(y_st/\tau_{B_s})}.
\end{equation}
The observables entering this expressions do not depend on the decay constant $F_{B_s}$, 
in contrast to the branching ratio in (\ref{BR-par}), and read as follows:
\begin{equation}
C_\lambda\equiv\frac{1-|\xi_\lambda|^2}{1+|\xi_\lambda|^2}
=-\eta_\lambda\left[\frac{2|PS|\cos(\varphi_P-\varphi_S)}{|P|^2+|S|^2}  \right]
\quad\stackrel{\rm SM}{\longrightarrow}\quad 0
\end{equation}
\begin{equation}
S_\lambda\equiv \frac{2\,\mbox{Im}\,\xi_\lambda}{1+|\xi_\lambda|^2}
=\frac{|P|^2\sin (2\varphi_P-\phi^{\rm NP}_s)-|S|^2\sin (2\varphi_S-\phi^{\rm NP}_s)}{|P|^2+|S|^2}
\quad \, \stackrel{\rm SM}{\longrightarrow}\quad 0
\end{equation}
\begin{equation}
{\cal A}_{\Delta\Gamma}^\lambda\equiv \frac{2\,\mbox{Re}\,\xi_\lambda}{1+|\xi_\lambda|^2}
=\frac{|P|^2\cos (2\varphi_P-\phi^{\rm NP}_s)-|S|^2\cos(2\varphi_S-\phi^{\rm NP}_s)}{|P|^2+|S|^2}
\hspace*{0.3truecm}\stackrel{\rm SM}{\longrightarrow}\hspace*{0.3truecm} 1,
\end{equation}
where also the corresponding SM values are given. The phase 
$\phi^{\rm NP}_s$ is the NP component of the $B^0_s$--$\bar B^0_s$ mixing phase
\begin{equation}
\phi_s=\phi_s^{\rm SM}+\phi_s^{\rm NP}=-2\lambda^2\eta+\phi_s^{\rm NP}.
\end{equation}
It should be noted that 
\begin{equation}
{\cal S}_{\mu\mu} \equiv S_\lambda \quad\mbox{and}\quad  
{\cal A}_{\Delta\Gamma}^{\mu\mu}\equiv {\cal A}_{\Delta\Gamma}^\lambda
\end{equation}
are {\it independent} of the muon helicity $\lambda$.  Neglecting the impact of 
$\Delta\Gamma_s$, CP asymmetries in $B_{s,d}\to \ell^+\ell^-$ decays were considered
for various NP scenarios in the previous literature.\cite{HL,DP,CKWW}

Since it is difficult to measure the muon helicity, we consider the rate
\begin{equation}
\Gamma(B_s^0(t)\to \mu^+\mu^-)\equiv \sum_{\lambda={\rm L,R}}
\Gamma(B_s^0(t)\to \mu^+_\lambda \mu^-_\lambda),
\end{equation}
and in analogy for the CP-conjugate process. The corresponding CP-violating 
rate asymmetry takes the form
\begin{equation}\label{S-asym}
\frac{\Gamma(B^0_s(t)\to \mu^+\mu^-)-
\Gamma(\bar B^0_s(t)\to \mu^+\mu^-)}{\Gamma(B^0_s(t)\to \mu^+\mu^-)+
\Gamma(\bar B^0_s(t)\to \mu^+\mu^-)}
=\frac{{\cal S}_{\mu\mu}\sin(\Delta M_st)}{\cosh(y_st/ \tau_{B_s}) + 
{\cal A}_{\Delta\Gamma}^{\mu\mu} \sinh(y_st/ \tau_{B_s})},
\end{equation}
where the $C_\lambda\propto \eta_\lambda$ terms  entering Eq.~(\ref{TD-CP}) 
cancel. It would be most interesting to measure this CP asymmetry as a non-zero 
value would signal CP-violating NP phases, as will be discussed in detail 
below. Unfortunately, this is challenging in view of the tiny branching ratio and as 
tagging and time information are required.

\subsection{Untagged Rate and Branching Ratio}
The first measurements of the $B_s\to \mu^+\mu^-$ channel discussed in Section~\ref{sec:intro}
concern the experimental branching ratio:
\begin{equation}
\overline{\rm BR}(B_{s}\to\mu^+\mu^-)
        \equiv \frac{1}{2}\int_0^\infty \langle \Gamma(B_s(t)\to \mu^+\mu^-)\rangle\, dt,
\end{equation}
i.e.\ the time-integrated untagged rate, with
\begin{eqnarray}
\lefteqn{\langle \Gamma(B_s(t)\to \mu^+\mu^-)\rangle\equiv
\Gamma(B^0_s(t)\to  \mu^+\mu^-)+ \Gamma(\bar B^0_s(t)\to  \mu^+\mu^-)}\nonumber\\
&&\propto e^{-t/\tau_{B_s}}\bigl[\cosh(y_st/ \tau_{B_s})+  {\cal A}_{\Delta\Gamma}^{\mu\mu}
\sinh(y_st/ \tau_{B_s})\bigr].
\end{eqnarray}
With the help of the relation
\begin{equation}
        {\rm BR}(B_s \to \mu^+\mu^-)= 
        \left[\frac{1-y_s^2}{1 + {\cal A}^{\mu\mu}_{\Delta\Gamma}\, y_s}\right]  \,
        \overline{\rm BR}(B_{s}\to\mu^+\mu^-),
\end{equation}
we may convert the experimental branching ratio into the theoretical branching ratio, which
refers to $t=0$. The observable ${\cal A}^{\mu\mu}_{\Delta\Gamma}$ depends on possible
NP contributions and is hence unknown, thereby satisfying 
${\cal A}^{\mu\mu}_{\Delta\Gamma}\in[-1,+1]$. Consequently, there are two options for
dealing with the theoretical interpretation of the measured, time-integrated branching ratio: 
\begin{itemize}
\item[(i)] Add an extra error to the experimental branching ratio: 
\begin{equation}
\Delta {\rm BR}(B_s \to \mu^+\mu^-)|_{y_s}=\pm y_s 
\overline{\rm BR}(B_{s}\to\mu^+\mu^-).
\end{equation}
\item[(ii)] Use the SM value ${\cal A}^{\mu\mu}_{\Delta\Gamma}|_{\rm SM}=+1$ to 
calculate a {\it new SM reference value}
for the comparison with the time-integrated experimental branching ratio 
$\overline{\rm BR}(B_{s}\to\mu^+\mu^-)$. To this end, $\mbox{BR}(B_s \to \mu^+\mu^-)_{\rm SM}$ 
has to be rescaled by the factor $1/(1-y_s)$, which results in
\begin{equation}
\overline{\rm BR}(B_{s}\to\mu^+\mu^-)_{\rm SM}= (3.65\pm0.23)\times 10^{-9}.
\end{equation}
\end{itemize}
Once more and more data are collected and also the decay time information for the 
untagged data sample is used, the effective $B_s\to \mu^+\mu^-$ lifetime 
\begin{equation}
\tau_{\mu\mu} \equiv \frac{\int_0^\infty t\,\langle \Gamma(B_s(t)\to \mu^+\mu^-)\rangle\, dt}
        {\int_0^\infty \langle \Gamma(B_s(t)\to \mu^+\mu^-)\rangle\, dt}
\end{equation}
can be determined, which allows the extraction of the observable 
\begin{equation}
{\cal A}^{\mu\mu}_{\Delta\Gamma}  = \frac{1}{y_s}\left[
\frac{(1-y_s^2)\tau_{\mu\mu}-(1+
 y_s^2)\tau_{B_s}}{2\tau_{B_s}-(1-y_s^2)\tau_{\mu\mu}}\right].
\end{equation}
The physics information encoded in $\tau_{\mu\mu}$ and ${\cal A}^{\mu\mu}_{\Delta\Gamma}$
is equivalent. Finally, using 
\begin{equation}
{\rm BR}\left(B_s \to \mu^+\mu^-\right)=\left[2 - \left(1-y_s^2\right)
\frac{\tau_{\mu\mu}}{\tau_{B_s}}\right]\overline{\rm BR}(B_{s}\to\mu^+\mu^-),
\end{equation}
which involves only measurable quantities on the right-hand side, 
the theoretical $B_s\to\mu^+\mu^-$
branching ratio can be extracted directly from the data.\cite{Bsmumu-paper}
Authors have started to include the effect of $\Delta\Gamma_s$ in analyses of constraints on 
NP which are implied by the measured $\overline{\rm BR}(\Bsmumu)$ 
in the recent literature (see, for instance, Refs.\ 29 and 34--40).

\section{Probing New Physics with $B^0_s\to\mu^+\mu^-$}	
In the following, we assume that the $B^0_s$--$\bar B^0_s$ mixing phase 
$\phi_s$ will be precisely known by the time the $B_s\to\mu^+\mu^-$ observables introduced
in the previous section can be measured,\cite{Rev,wouter} 
which will allow the determination of the NP phase $\phi_s^{\rm NP}$.

\begin{figure}
\centerline{\includegraphics[width=6.4cm]{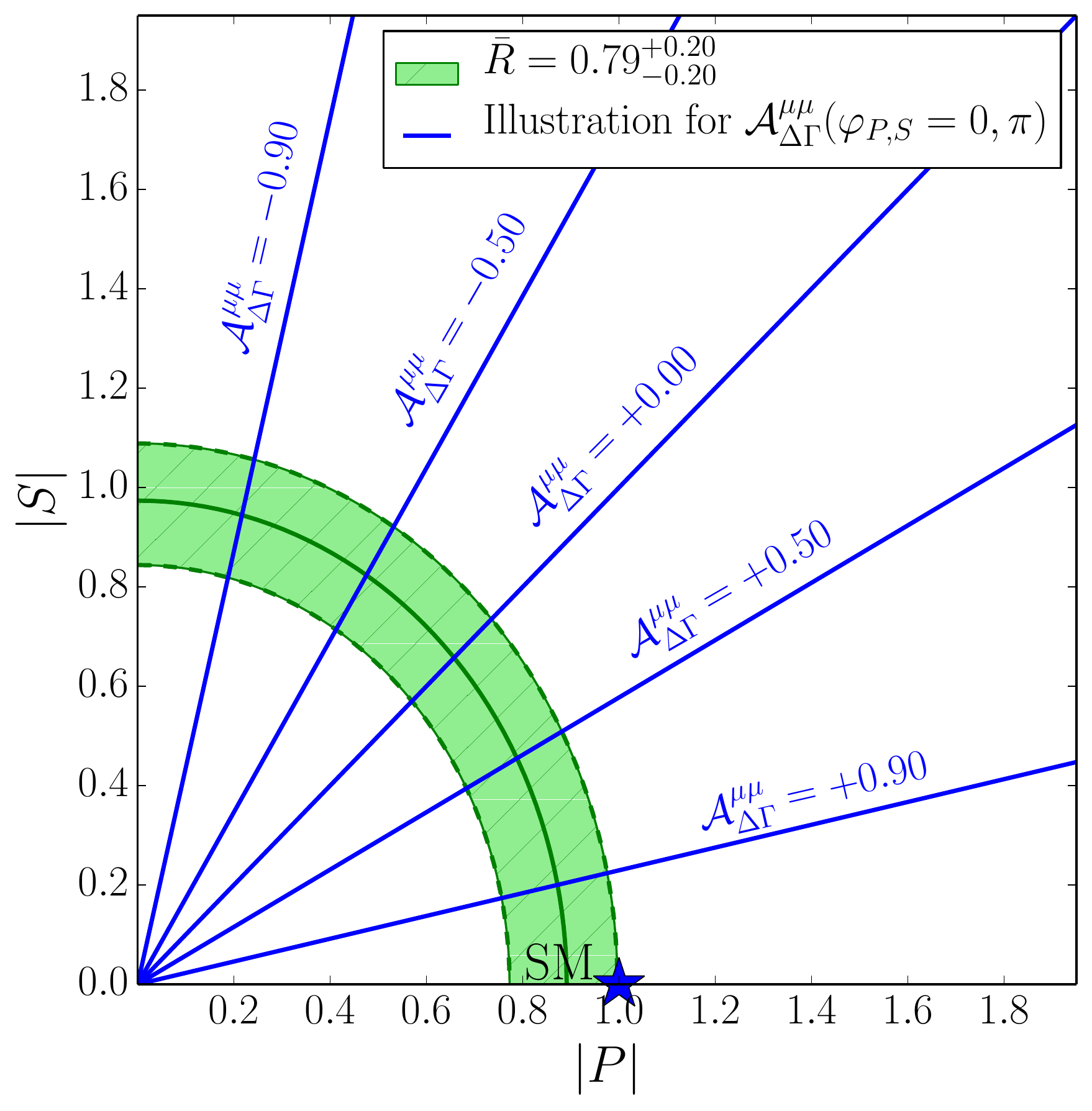}}
\caption{The situation in the $P$--$S$ plane, where the circular band corresponds to the
allowed region following from (\ref{R-value}), and contours for various values of the
observable ${\cal A}_{\Delta\Gamma}^{\mu\mu}$ are shown, assuming 
no CP-violating NP phases $\varphi_P$ and $\varphi_S$.\label{fig-PS}}
\end{figure}

In order to perform a test of the SM with the measurement of the $B_s\to\mu^+\mu^-$
branching ratio, it is useful to introduce the following ratio:\cite{Bsmumu-paper,BFGK}
\begin{eqnarray}
\lefteqn{\overline{R} \equiv \frac{\overline{\rm BR}(\Bsmumu)}{\overline{\rm BR}(\Bsmumu)_{\rm SM}}
        = \left[\frac{1+{\cal A}^{\mu\mu}_{\Delta\Gamma}\,y_s}{1+y_s} \right]  (|P|^2 + |S|^2)}\nonumber\\
&&= \left[\frac{1+y_s\cos(2\varphi_P-\phi_s^{\rm NP})}{1+y_s} \right] |P|^2 + \left[\frac{1-y_s\cos(2\varphi_S-\phi_s^{\rm NP})}{1+y_s} \right] |S|^2.\label{Rbar-def}
\end{eqnarray}
The current situation corresponds to 
\begin{equation}\label{R-value}
\overline{R} = 0.79 \pm 0.20. 
\end{equation}
Looking at Eq.~(\ref{Rbar-def}), we observe that $\overline{R}$ does not 
allow a separation of the $P$ and $S$ contributions. Sizable NP contributions 
could be present in $B_s\to\mu^+\mu^-$, even if $\overline{R}$ is measured with a value
close to $\overline{R}_{\rm SM}=1$.

This situation can be resolved with the help of additional information, which is offered by the
measurement of the effective lifetime $\tau_{\mu\mu}$ or -- equivalently -- 
the observable ${\cal A}^{\mu\mu}_{\Delta\Gamma}$: 
\begin{equation}
|S|=|P|\sqrt{\frac{\cos(2\varphi_P-\phi^{\rm NP}_s)
-{\cal A}_{\Delta\Gamma}^{\mu\mu}}{\cos(2\varphi_S 
-\phi^{\rm NP}_s)+{\cal A}_{\Delta\Gamma}^{\mu\mu}}}.
\end{equation}
In Fig.~\ref{fig-PS}, the current constraints in the $|P|$--$|S|$ plane are shown with an
illustration of the contours following from a future measurement of 
${\cal A}_{\Delta\Gamma}^{\mu\mu}$. In the calculation
of the latter curves, it was assumed that there are no CP-violating NP phases
$\varphi_P$ and $\varphi_S$, as is the case, for instance, in NP models exhibiting 
MFV without flavour-blind phases. 

In Ref.~\refcite{BFGK}, detailed discussions of the regions in the 
$\overline{R}$--${\cal A}_{\Delta\Gamma}^{\mu\mu}$ plane that are allowed by current
data are given for a variety of scenarios:
\begin{itemize}
\item $P=1+\tilde P$ ($\tilde P$ free) and $S=0$: the deviation of 
${\cal A}_{\Delta\Gamma}^{\mu\mu}$ from its SM value $+1$ requires 
CP-violating NP phases.  [Examples: CMFV, LHT, 4G, RSc, $Z'$]
\item $P=1$ and $S$ free: 
${\cal A}_{\Delta\Gamma}^{\mu\mu}$ may differ from its SM value $+1$ without new 
CP-violating phases, with $\overline{R} \geq 1$. The experimental data have 
already quite some impact in this scenario.
[Example: 2HDM (scalar $H^0$ dominance)]
\item $P\pm S=1$: the 
full range of ${\cal A}_{\Delta\Gamma}^{\mu\mu}$ can be accessed without new 
CP-violating phases, and the following lower bound arises:
\begin{equation}
\overline{\rm BR}(\Bsmumu) \geq \frac{1}{2}\left( 1-y_s \right)
\overline{\rm BR}(\Bsmumu)_{\rm SM}.
\end{equation}
[Example: decoupled 2HDM/MSSM ($M_{H^0}\approx M_{A^0}\gg M_{h^0}$)]
\end{itemize}

\begin{figure}
\centerline{\includegraphics[width=8.0cm]{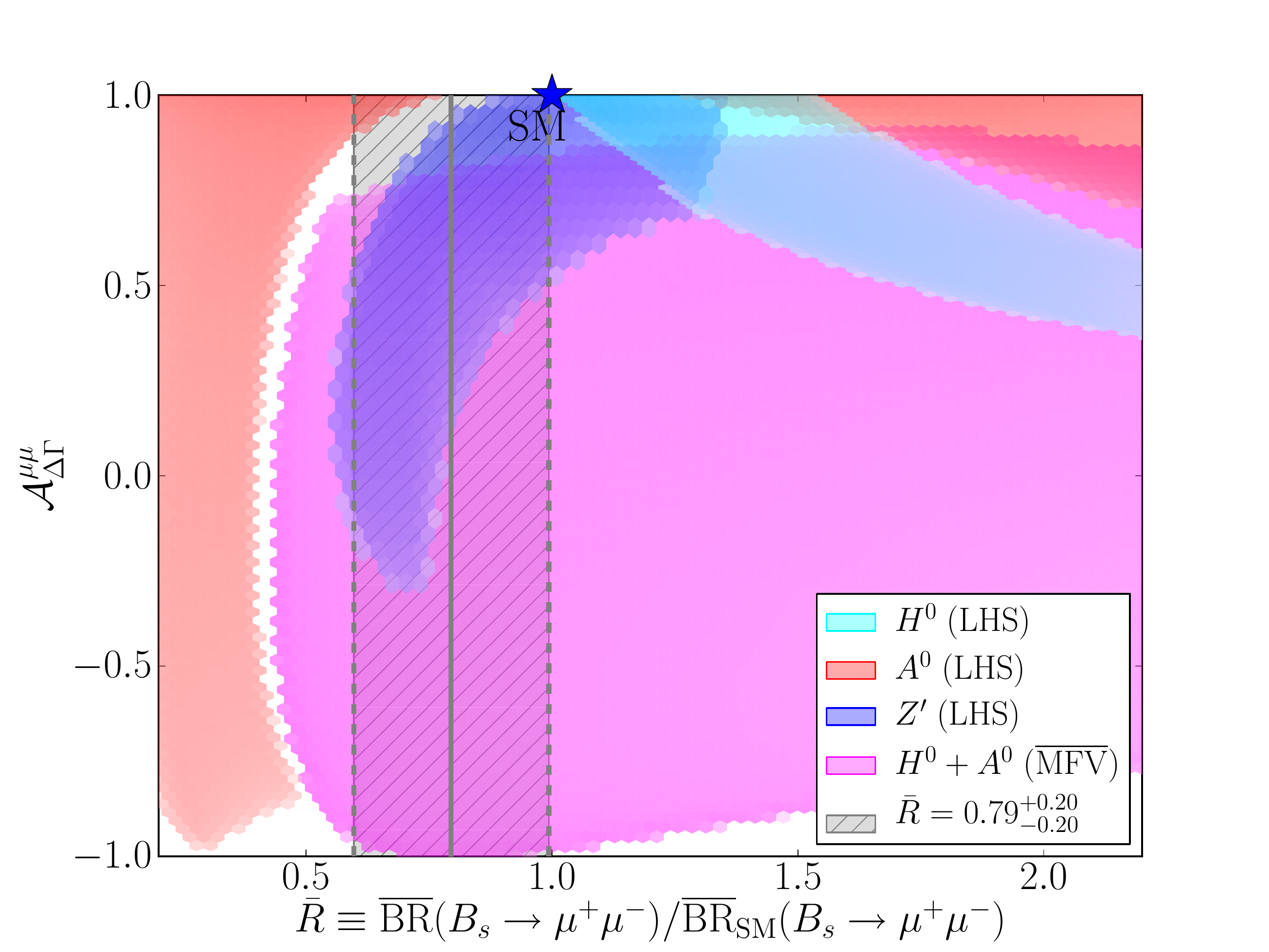}}
\caption{Allowed regions in the $\overline{R}$--${\cal A}_{\Delta\Gamma}^{\mu\mu}$ plane
for various NP scenarios.\cite{BFGK}.\label{fig-RADG-const}}
\end{figure}

\begin{figure}
\centerline{\includegraphics[width=8.0cm]{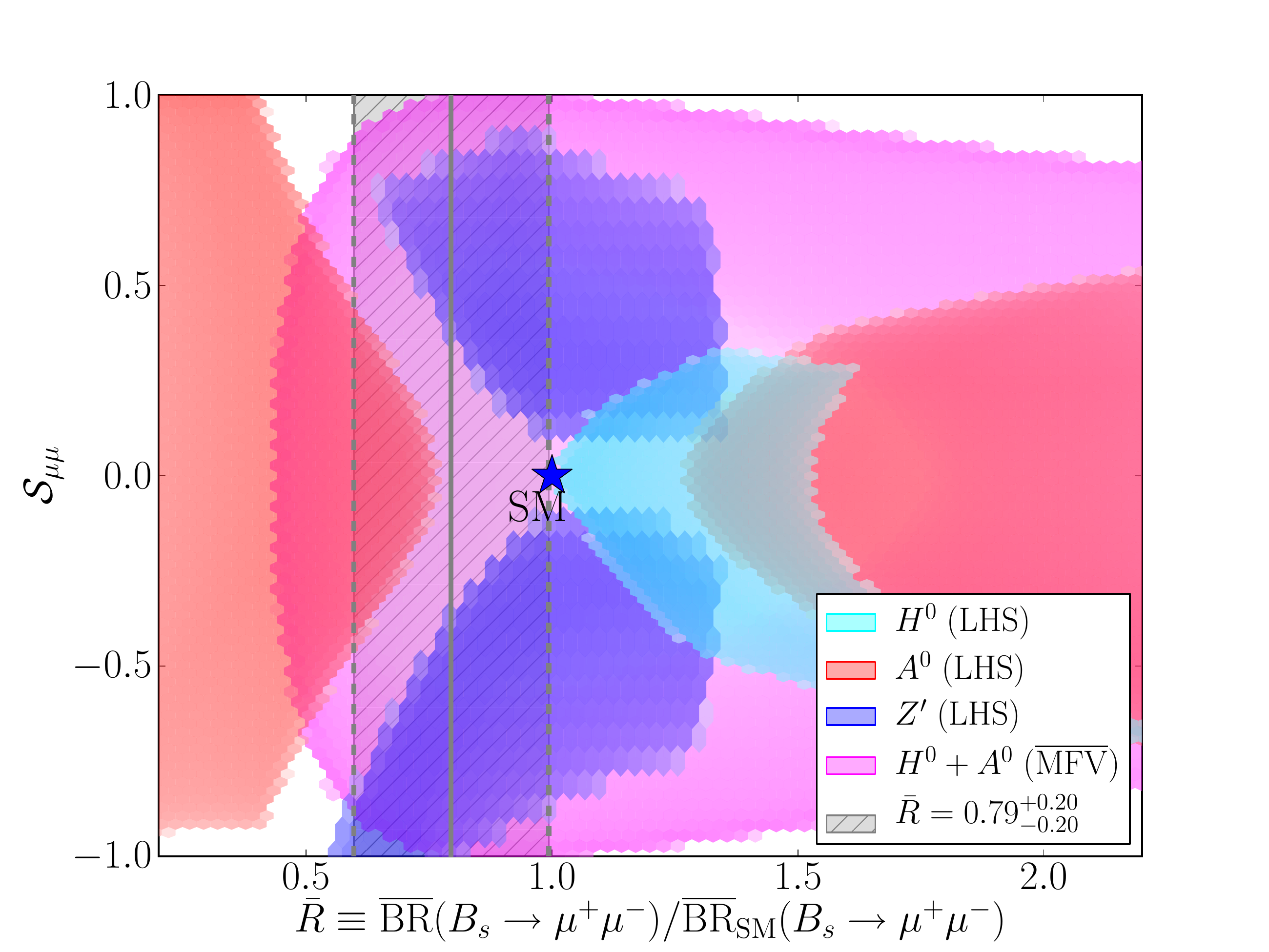}}
\caption{Allowed regions in the $\overline{R}$--${\cal S}_{\mu\mu}$ plane
for various NP scenarios.\cite{BFGK}.\label{fig-RS-const}}
\end{figure}

In addition to the discussion of these general scenarios, also detailed analyses within 
specific NP models were performed in Ref.~\refcite{BFGK}:
\begin{itemize}
\item Tree-level neutral gauge boson exchange, characterised by 
\begin{equation}
 \mathcal{L}_{{\rm FCNC}}(Z')=\left[\Delta_L^{sb}(Z')(\bar s \gamma_\mu P_L b)+
                      \Delta_R^{sb}(Z')(\bar s \gamma_\mu P_R b)\right] Z^{'\mu}.                 
\end{equation}
Various scenarios can be distinguished: left-handed scheme (LHS) with complex 
$\Delta_L^{bs}\not=0$  and $\Delta_R^{bs}=0$, right-handed scheme (RHS) with 
complex $\Delta_R^{bs}\not=0$  and $\Delta_L^{bs}=0$, 
left-right symmetric scheme (LRS) with complex
$\Delta_L^{bs}=\Delta_R^{bs}\not=0$, left-right asymmetric scheme (ALRS) 
with complex $\Delta_L^{bs}=-\Delta_R^{bs}\not=0$.
\item Tree-level neutral (pseudo)scalar exchange:
\begin{equation}
\mathcal{L}_{\rm FCNC}(H)=\left[\Delta_L^{sb}(H)(\bar s P_L b)+
                      \Delta_R^{sb}(H)(\bar s P_R b)\right] H.
\end{equation}
\item Tree-level neutral scalar+pseudoscalar exchange:
\begin{eqnarray}
    {\cal L}_{\rm FCNC}(H^0,A^0)
    =& \left[\Delta_L^{sb}(H^0)(\bar s P_L b)+
                      \Delta_R^{sb}(H^0)(\bar s P_R b)\right] H^0 \nonumber\\
    &+ \left[\Delta_L^{sb}(A^0)(\bar s P_L b)+
                      \Delta_R^{sb}(A^0)(\bar s P_R b)\right] A^0 .
                    \end{eqnarray}
\end{itemize}
In the exploration of the allowed regions in observable space, it is important to take
also experimental constraints on $B^0_s$--$\bar B^0_s$ mixing and other rare decays
into account. 

In Fig.~\ref{fig-RADG-const}, the
constraints in the $\overline{R}$--${\cal A}^{\mu\mu}_{\Delta\Gamma}$ plane are shown, involving
only {\it untagged} observables of the $B_s\to\mu^+\mu^-$ channel. The allowed regions in this 
figure are complemented by those in Fig.~\ref{fig-RS-const}, showing the situation in the
$\overline{R}$--${\cal S}_{\mu\mu}$ plane. Here {\it tagging} is required for the measurement of the
CP asymmetry ${\cal S}_{\mu\mu}$, as can be seen in Eq.~(\ref{S-asym}). It is interesting to
note the relation 
\begin{equation}
        | {\cal S}_{\mu\mu}|^2 + |{\cal A}^{\mu\mu}_{\Delta\Gamma}|^2 
        = 1 - \left[\frac{2|PS|\cos(\varphi_P - \varphi_S)}{|P|^2 + |S|^2}\right]^2,
\end{equation}
which illustrates the sensitivity to CP-violating NP phases.

\section{Summary and Outlook}	
We live in exciting times for studies of leptonic rare $B_{(s)}$-meson decays. Concerning 
$B_d\to\mu^+\mu^-$, the average of the CMS and LHCb results for the branching ratio
is given by $\left(3.6^{+1.6}_{-1.4}\right)\times 10^{-10}$ while the SM prediction reads
$(1.06\pm0.09)\times 10^{-10}$. The large central value of the 
experimental average would immediately imply NP and would rule out models with MFV. But
the large error does unfortunately not allow us to draw conclusions at this point. It will be
very interesting to monitor the future evolution of the data. 

Concerning $B_s\to \mu^+\mu^-$, there is finally evidence for this channel in the data of the
CMS and LHCb collaborations.  The corresponding LHC average for the measured branching 
ratio is given by $\overline{\rm BR}(B_{s}\to\mu^+\mu^-) = (2.9\pm0.7 ) \times 10^{-9}$. 
This result falls well into the SM regime although the error is still sizeable. 

The interpretation of this measurement is affected by a -- seemingly -- unrelated topic, which is
the recent development that a non-vanishing value of the $B_s$ decay width difference
$\Delta\Gamma_s$ was established by the LHCb collaboration. In view of this result, 
special care has to be taken when dealing with branching ratios of $B_s$ decays, distinguishing 
in particular between the ``experimental" and ``theoretical" branching ratios. 
Apart from this complication, $\Delta\Gamma_s$ offers new observables which are 
encoded in the effective lifetimes of $B_s$ decays. The most relevant application 
concerns the search for NP with $B_s\to\mu^+\mu^-$. 

The SM reference value for the comparison with the time-integrated experimental branching ratio 
including the $\Delta\Gamma_s$ effects is given as follows:
\begin{equation}
\overline{\rm BR}(B_{s}\to\mu^+\mu^-)_{\rm SM}= (3.65\pm0.23)\times 10^{-9},
\end{equation}
using the most recent theoretical analysis of Refs.~\refcite{Theo-1,Theo-2,Theo-3}.

Apart from a more precise measurement of
$\overline{\rm BR}(B_{s}\to\mu^+\mu^-)$, the next conceptual step is the analysis of the
{\it time-dependent} untagged $B_s\to\mu^+\mu^-$ rate and the determination of the
effective lifetime $\tau_{\mu\mu}$. The sizeable $\Delta\Gamma_s$ offers access 
to ${\cal A}^{\mu\mu}_{\Delta\Gamma}$, which is a new {\it theoretically clean} observable 
to search for NP. In contrast to the branching ratio, the dependence 
on the $B_s$ decay constant $F_{B_s}$ cancels out. Interestingly, 
${\cal A}^{\mu\mu}_{\Delta\Gamma}$ may reveal NP effects even if the branching 
ratio is found close to the SM prediction.
Using in addition tagging information to distinguish between initially present $B^0_s$ or
$\bar B^0_s$ mesons, the CP asymmetry ${\cal S}_{\mu\mu}$ can be extracted from the
time-dependent rates. 

Thanks to still largely unconstrained (pseudo-)scalar operators 
$O_{(P)S}$, $O_{(P)S}'$ there is still sizeable space for NP to manifest itself 
in these observables. Correlations between $\overline{R}$, ${\cal A}^{\mu\mu}_{\Delta\Gamma}$ 
and ${\cal S}_{\mu\mu}$ allow us to distinguish between different NP scenarios, with 
their specific effective operators and CP-violating phases. A particularly exciting scenario
would be the detection of sources of CP violation originating from physics beyond the SM. 
The new $B_s\to \mu^+\mu^-$ observables offer 
new studies for the LHC upgrade physics programme and may allow us 
to eventually reveal NP in $B_s\to\mu^+\mu^-$, one of the rarest and most
fascinating decays Nature has to offer.

\section*{Acknowledgments}
I would like to thank my PhD students and colleagues for the enjoyable collaboration on topics 
discussed above, and Rob Knegjens for numerical updates. I am very grateful to Harald Fritzsch
and his co-organizers for inviting me to this excellent and most enjoyable workshop on Flavor Physics and Mass Generation and for their kind hospitality in Singapore.
%
%
%

%
%
%

\begin{thebibliography}{00}  

\bibitem{BGGI}A.~J.~Buras, J.~Girrbach, D.~Guadagnoli and G.~Isidori,
  Eur.\ Phys.\ J.\ C {\bf 72}, 2172 (2012)
  [arXiv:1208.0934 [hep-ph]].
  
\bibitem{BFGK}A.~J.~Buras, R.~Fleischer, J.~Girrbach and R.~Knegjens,
  JHEP {\bf 1307}, 77 (2013)
  [arXiv:1303.3820 [hep-ph]].

\bibitem{Lattice} R.~J.~Dowdall {\it et al.}  [HPQCD Collaboration],
  Phys.\ Rev.\ Lett.\  {\bf 110}, 222003 (2013)
  [arXiv:1302.2644 [hep-lat]].
  
\bibitem{HFAG}Y.~Amhis {\it et al.}  [Heavy Flavor Averaging Group Collaboration],
  {\it Averages of B-Hadron, C-Hadron, and tau-lepton properties as of early 2012},
  arXiv:1207.1158 [hep-ex]; 
for updates, see http://www.slac.stanford.edu/xorg/hfag/
  
\bibitem{Theo-1}C.~Bobeth, M.~Gorbahn, T.~Hermann, M.~Misiak, E.~Stamou and M.~Steinhauser,
  Phys.\ Rev.\ Lett.\  {\bf 112}, 101801 (2014)
  [arXiv:1311.0903 [hep-ph]].

\bibitem{Theo-2}C.~Bobeth, M.~Gorbahn and E.~Stamou,
  Phys.\ Rev.\ D {\bf 89}, 034023 (2014)
  [arXiv:1311.1348 [hep-ph]].

\bibitem{Theo-3}T.~Hermann, M.~Misiak and M.~Steinhauser,
  JHEP {\bf 1312}, 097 (2013)
  [arXiv:1311.1347 [hep-ph]].
  
\bibitem{straub}D.~M.~Straub,
  arXiv:1012.3893 [hep-ph].

\bibitem{BuGir-12}A.~J.~Buras and J.~Girrbach,
  Acta Phys.\ Polon.\ B {\bf 43}, 1427 (2012)
  [arXiv:1204.5064 [hep-ph]].

\bibitem{CDF-bound}T.~Aaltonen {\it et al.}  [CDF Collaboration],
  Phys.\ Rev.\ D {\bf 87}, 072003 (2013)
  [arXiv:1301.7048 [hep-ex]].

\bibitem{D0-bound}V.~M.~Abazov {\it et al.}  [D0 Collaboration],
  Phys.\ Rev.\ D {\bf 87},  072006 (2013)
  [arXiv:1301.4507 [hep-ex]].
  
\bibitem{ATLAS-mumu}G.~Aad {\it et al.}  [ATLAS Collaboration],
  Phys.\ Lett.\ B {\bf 713}, 387 (2012)
  [arXiv:1204.0735 [hep-ex]];
  ATLAS-CONF-2013-076 (2013).

\bibitem{CMS-mumu}S.~Chatrchyan {\it et al.}  [CMS Collaboration],
  Phys.\ Rev.\ Lett.\  {\bf 111}, 101804 (2013)
  [arXiv:1307.5025 [hep-ex]].

\bibitem{LHCb-mumu}R.~Aaij {\it et al.}  [LHCb Collaboration],
  Phys.\ Rev.\ Lett.\  {\bf 111}, 101805 (2013)
  [arXiv:1307.5024 [hep-ex]].

\bibitem{LHC-Bsmumu-average}The CMS and LHCb Collaborations, 
CMS-PAS-BPH-13-007,  LHCb-CONF-2013-012 (2013).

\bibitem{FST}R.~Fleischer, N.~Serra and N.~Tuning,
  Phys.\ Rev.\ D {\bf 82}, 034038 (2010)
  [arXiv:1004.3982 [hep-ph]];
  Phys.\ Rev.\ D {\bf 83}, 014017 (2011)
  [arXiv:1012.2784 [hep-ph]].
  
\bibitem{FF-lat}J.~A.~Bailey, A.~Bazavov, C.~Bernard, C.~M.~Bouchard, C.~DeTar, D.~Du, A.~X.~El-Khadra and J.~Foley {\it et al.},
  Phys.\ Rev.\ D {\bf 85}, 114502 (2012)
  [Erratum-ibid.\ D {\bf 86}, 039904 (2012)]
  [arXiv:1202.6346 [hep-lat]].
 
 \bibitem{RF-rev}R.~Fleischer,
  Phys.\ Rept.\  {\bf 370}, 537 (2002)
  [hep-ph/0207108].
 
 \bibitem{Lenz}A.~Lenz,
  {\it Theoretical update of $B$-Mixing and Lifetimes},
  arXiv:1205.1444 [hep-ph].
   
\bibitem{LHCb-DGs}R. Aaij {\it et al.}  [LHCb Collaboration],
  Phys.\ Rev.\ D {\bf 87}, 112010 (2013)
  [arXiv:1304.2600 [hep-ex]].

\bibitem{BR-paper}K.~De Bruyn, R.~Fleischer, R.~Knegjens, P.~Koppenburg, M.~Merk and N.~Tuning,
  Phys.\ Rev.\ D {\bf 86}, 014027 (2012)
  [arXiv:1204.1735 [hep-ph]].
 
\bibitem{Bsmumu-paper}K.~De Bruyn, R.~Fleischer, R.~Knegjens, P.~Koppenburg, M.~Merk, A.~Pellegrino and N.~Tuning,
  Phys.\ Rev.\ Lett.\  {\bf 109}, 041801 (2012)
  [arXiv:1204.1737 [hep-ph]].

\bibitem{DDF}I.~Dunietz, R.~Fleischer and U.~Nierste,
  Phys.\ Rev.\ D {\bf 63}, 114015 (2001)
  [hep-ph/0012219].

\bibitem{RF-JpsiK}R.~Fleischer,
  Eur.\ Phys.\ J.\ C {\bf 10}, 299 (1999)
  [hep-ph/9903455].

\bibitem{RF-KK}R.~Fleischer,
  Phys.\ Lett.\ B {\bf 459}, 306 (1999)
  [hep-ph/9903456].

\bibitem{FFM}S.~Faller, R.~Fleischer and T.~Mannel,
  Phys.\ Rev.\ D {\bf 79}, 014005 (2009)
  [arXiv:0810.4248 [hep-ph]].
  
  \bibitem{LHCb-BsKastKast}R.~Aaij {\it et al.}  [LHCb Collaboration],
  Phys.\ Lett.\ B {\bf 709}, 50 (2012) [arXiv:1111.4183 [hep-ex]].

\bibitem{DGMV}S.~Descotes-Genon, J.~Matias and J.~Virto,
  Phys.\ Rev.\ D {\bf 85},  034010 (2012)  [arXiv:1111.4882 [hep-ph]].
  
 \bibitem{AS}W.~Altmannshofer and D.~M.~Straub,
  JHEP {\bf 1208}, 121 (2012)
  [arXiv:1206.0273 [hep-ph]].
  
 \bibitem{WA}W.~Altmannshofer,
  PoS Beauty {\bf 2013}, 024 (2013)
  [arXiv:1306.0022 [hep-ph]].
  
\bibitem{HL} C.-S.~Huang and W.~Liao,
  Phys.\ Lett.\ B {\bf 525},  107  (2002) [hep-ph/0011089];
  Phys.\ Lett.\ B {\bf 538}, 301 (2002)
  [hep-ph/0201121].
  
\bibitem{DP}A.~Dedes and A.~Pilaftsis,
  Phys.\ Rev.\ D {\bf 67},   015012  (2003) [hep-ph/0209306].

\bibitem{CKWW}P.~H.~Chankowski, J.~Kalinowski, Z.~Was and M.~Worek,
  Nucl.\ Phys.\ B {\bf 713},  555   (2005) [hep-ph/0412253].
   
\bibitem{LLP}X.~-Q.~Li, J.~Lu and A.~Pich,
  JHEP {\bf 1406}, 022 (2014)
  [arXiv:1404.5865 [hep-ph]].
  
\bibitem{ACSY}W.~Altmannshofer, M.~Carena, N.~R.~Shah and F.~Yu,
  JHEP {\bf 1301},  160 (2013)
  [arXiv:1211.1976 [hep-ph]].
 
\bibitem{BFG}A.~J.~Buras, F.~De Fazio and J.~Girrbach,
  JHEP {\bf 1302},  116  (2013)  [arXiv:1211.1896 [hep-ph]].
    
\bibitem{BCCDDEFH}O.~Buchmueller, R.~Cavanaugh, M.~Citron, A.~De Roeck, M.~J.~Dolan, J.~R.~Ellis, H.~Flacher and S.~Heinemeyer {\it et al.},
  Eur.\ Phys.\ J.\ C {\bf 72}, 2243 (2012)
  [arXiv:1207.7315].
  
\bibitem{TM}T.~Hurth and F.~Mahmoudi,
 Nucl.\ Phys.\ B {\bf 865} (2012) 461 [arXiv:1207.0688 [hep-ph]].
  
\bibitem{BKMS}D.~Becirevic, N.~Kosnik, F.~Mescia and E.~Schneider,
  Phys.\ Rev.\ D {\bf 86}, 034034 (2012)
  [arXiv:1205.5811 [hep-ph]].
  
\bibitem{MNO}F.~Mahmoudi, S.~Neshatpour and J.~Orloff,
  JHEP {\bf 1208}, 092 (2012)
  [arXiv:1205.1845 [hep-ph]].

\bibitem{Rev}G.~Borissov, R.~Fleischer and M.-H.~Schune,
  Ann.\ Rev.\ Nucl.\ Part.\ Sci.\  {\bf 63}, 205 (2013)
  [arXiv:1303.5575 [hep-ph]].

\bibitem{wouter}W.~Hulsbergen,
  Mod.\ Phys.\ Lett.\ A {\bf 28}, 1330023 (2013)
  [arXiv:1306.6474 [hep-ph]].

%
%
%
\end{thebibliography}
\end{document}